\documentclass[showpacs,twocolumn,showkeys,amsmath,amssymb,pra,superscriptaddress,nofootinbib]{revtex4-1} 

\usepackage{amsmath}
\usepackage{color}
\usepackage{graphicx}   
\usepackage{dcolumn}    
\usepackage{bm}         
\begin{document} 
 
\title[]{Beyond mean-field behavior of large Bose-Einstein condensates in double-well potentials}
\author{Bettina Gertjerenken}
\email{b.gertjerenken@uni-oldenburg.de}
\affiliation{Institut f\"ur Physik, Carl von Ossietzky Universit\"at, D-26111 Oldenburg, Germany}
\author{Christoph Weiss}
\affiliation{Institut f\"ur Physik, Carl von Ossietzky Universit\"at, D-26111 Oldenburg, Germany}
\affiliation{Joint Quantum Centre  (JQC) Durham--Newcastle, Department of Physics, Durham University, Durham DH1 3LE, United Kingdom}

\keywords{{double well}, mesoscopic entanglement, Bose-Einstein condensation, mean-field limit, {separatrix}}
                  
\date{\today}
 
\begin{abstract}
For the dynamics of Bose-Einstein condensates (BECs), differences between mean-field (Gross-Pitaevskii) physics and $N$-particle quantum physics often disappear if the BEC becomes larger and larger. In particular, the timescale for which both dynamics agree should thus become larger if the particle number increases. For BECs in a double-well potential, we find both examples for which this is the case and examples for which differences remain even for huge BECs on experimentally realistic short timescales. By using a combination of numerical and analytical methods, we show that the differences remain visible on the level of expectation values even beyond the largest possible numbers realized experimentally for BECs with ultracold atoms.
\end{abstract}

\pacs{03.75.Gg,05.45.Mt,03.75.Lm}

\maketitle


\section{Introduction}
A widely used approach to describe both dynamics and ground-state properties of Bose-Einstein condensates (BEC)~\cite{BlochEtAl08} is the mean-field description via the  Gross-Pitaevskii equation~(GPE)~\cite{Pitaevskii61,Gross61,DalfovoEtAl99}. Within this mean-field approach, the BEC is characterized by the single particle density $|\Psi_0\left(\mathbf{r},t\right)|^2$; the time-dependence is given by
\begin{align}
{i}\hbar \frac{\partial}{\partial t}\Psi_0\left(\mathbf{r},t\right) =& \left[-\frac{\hbar^2\nabla^2}{2m}+V_{\mathrm{ext}}\left(\mathbf{r}\right) \right]\Psi_0\left(\mathbf{r},t\right)\nonumber \\
&+ (N-1)g|\Psi_0\left(\mathbf{r},t\right)|^2 \Psi_0\left(\mathbf{r},t\right),
\end{align}
where $V_{\mathrm{ext}}\left(\mathbf{r}\right)$ is an external potential, $g=\frac{4\pi \hbar^2a_{\rm{s}}}{m}$ an interaction parameter depending on the $s$-wave scattering length~$a_{\rm{s}}$ and $N$ is the particle number. The wave function $\Psi_0$ is normalized to one. The GPE has been used to describe topics as diverse as double-well potentials~\cite{SmerziEtAl97}, solitons~\cite{KivsharMalomed1989} or vortices~\cite{FetterSvidzinsky2001}.

In general, such a mean-field description might be expected to become better for larger particle-numbers $N$. In the mean-field limit~\cite{LiebEtAl00}
\begin{equation}
N\to\infty,\quad g\to 0,\quad \textrm{such that}\quad gN=\rm const,
\label{eq:meanfieldlimit}
\end{equation}
there even are cases for which it is possible to show that the GPE gives the correct ground-state energy~\cite{LiebEtAl00}. Accurate descriptions of ground state properties are also found in~\cite{LiebSeiringerEtAl03,LiebSeiringerEtAl09}. In~\cite{ErdoesEtAl07} dynamics of initially trapped Bose gases are investigated and it is proven that under certain conditions on the interaction potential and the initial state the time-evolution is correctly described by the GPE.

Nonetheless, noticeable differences between mean-field and $N$-particle dynamics exist. One example is the collapse and revival phenomenon~\cite{GreinerEtAl02b}: what appears to look like a classical damping can, in fact be followed by  at least a partial revival (cf.~\cite{HolthausStenholm01,Ziegler2011}). In general, situations with important quantum correlations, where a mean-field approach is no longer adequate, are in the focus of current research, e.~g. many-particle entanglement~\cite{PezzeSmerzi2009}, the experimental realization of entangled squeezed states~\cite{EsteveEtAl08}, mesoscopic quantum superpositions~\cite{PiazzaEtAl2008,CarrEtAl10,GertjerenkenEtAl12,DellAnna2012,TichyEtAl2013}, and mean-field chaos~\cite{GertjerenkenEtAl10,BrezinovaEtAl11}. In order to estimate timescales on which the mean-field dynamics still agrees with $N$-particle quantum dynamics, classical field methods can be used to approximate the quantum dynamics by averaging over mean-field solutions~\cite{UtermannEtAl94,SinatraEtAl2002,StrzysEtAl08,WeissTeichmann08,Chuchem10,BieniasEtAl2011,SimonStrunz12,KhripkovEtAl13,GertjerenkenEtAl2013} thus mimicking quantum uncertainties that disappear in the mean-field limit~(\ref{eq:meanfieldlimit}) but will always be present for finite particle numbers.

In order to investigate the differences between mean-field dynamics and quantum dynamics on the $N$-particle level in more detail, a BEC in a double-well potential is an ideal system~\cite{MilburnEtAl97,SmerziEtAl97,CastinDalibard97,RaghavanEtAl99b,AlbiezEtAl05,TeichmannEtAl06,LesanovskyEtAl06,PiazzaEtAl2008,LeeEtAl08,SakmannEtAl09,ZiboldEtAl10,MazzarellaEtAl2011}. While differences between mean-field dynamics and  $N$-particle quantum dynamics have been observed for small BECs~\cite{RaghavanEtAl99b,MilburnEtAl97}, it would be tempting to assume that Eq.~(\ref{eq:meanfieldlimit}) implies that those differences disappear if one simply chooses (experimentally realistic) large BECs.

While we do find cases for which this assumption is indeed correct [the ``quantum break time'' for which mean-field and $N$-particle quantum dynamics agree diverges in the mean-field limit~(\ref{eq:meanfieldlimit})], we also identify situations for which even for huge BECs this limit is not yet reached. Thus, it also is not reached for (experimentally realistic) large BECs.

The article is structured as follows. In Sec.~\ref{sec:model} the model system is introduced. For low interactions, Sec.~\ref{sec:Tmf} derives an ($N$-dependent) timescale on which $N$-particle dynamics and mean-field dynamics agree. While Sec.~\ref{sec:mfandmanybody} focuses on parameters with a good agreement of mean-field and $N$-particle results already for comparatively low particle numbers, in Sec.~\ref{sec:deviation} beyond mean-field behavior is discussed for very large condensates. Section~\ref{sec:conclusion} concludes the article.

\section{Model}\label{sec:model}
A BEC in a double well can be described with a model originally developed in nuclear physics~\cite{LipkinEtAl65}: a many-particle Hamiltonian in two-mode approximation~\cite{MilburnEtAl97}, 
\begin{eqnarray}
\label{eq:H}
\hat{H} &=& -\frac{\hbar\Omega}2\left(\hat{a}_1^{\dag}\hat{a}_2^{\phantom\dag} +\hat{a}_2^{\dag}\hat{a}_1^{\phantom\dag}\right) + \hbar\kappa\left(\hat{a}_1^{\dag}\hat{a}_1^{\dag}\hat{a}_1^{\phantom\dag}\hat{a}_1^{\phantom\dag}+\hat{a}_2^{\dag}\hat{a}_2^{\dag}\hat{a}_2^{\phantom\dag}\hat{a}_2^{\phantom\dag}\right)\nonumber\\
&+&\hbar\big(\mu_0+\mu_1\sin(\omega t)\big)\left(\hat{a}_2^{\dag}\hat{a}_2^{\phantom\dag}-\hat{a}_1^{\dag}\hat{a}_1^{\phantom\dag}\right)\;,
\end{eqnarray}
where the operator $\hat{a}^{(\dag)}_j$  annihilates (creates) a boson in well~$j$; $\hbar\Omega$ is the tunneling splitting, $2\hbar\mu_0$ is the tilt between well~1 and well~2 and $\hbar\mu_1$ is the driving amplitude. The interaction energy of a pair of particles in the same well is denoted by $2\hbar\kappa$.

The dynamics of a BEC in a double-well potential can be conveniently described using angular momentum operators:
\begin{eqnarray}\label{eq:angularmom}
\hat{J}_x & = & \frac{1}{2}\left(\hat{a}^{\phantom{\dagger}}_1\hat{a}_2^{\dagger} + \hat{a}_1^{\dagger}\hat{a}^{\phantom{\dagger}}_2\right),\nonumber \\
\hat{J}_y & = & -\frac{\mathrm{i}}{2}\left(\hat{a}^{\phantom{\dagger}}_1\hat{a}_2^{\dagger} - \hat{a}_1^{\dagger}\hat{a}^{\phantom{\dagger}}_2\right),\nonumber \\
\hat{J}_z & = & \frac{1}{2}\left( \hat{a}_1^{\dagger}\hat{a}^{\phantom{\dagger}}_1 - \hat{a}_2^{\dagger}\hat{a}^{\phantom{\dagger}}_2 \right).
\end{eqnarray}
The operator $J_z$ corresponds to the particle number difference between the two wells.

The Gross-Pitaevskii dynamics can be mapped to that of a nonrigid
pendulum~\cite{SmerziEtAl97}. Including the term  describing the  periodic shaking, the 
classical Hamiltonian is given by:
\begin{eqnarray}
\label{eq:mean}
H_{\rm mf}& = &\frac{N\kappa}{\Omega}z^2-\sqrt{1-z^2}\cos(\phi)\nonumber\\
&-&2z\left(\frac{\mu_0}{\Omega}+
\frac{\mu_1}{\Omega}\sin\left({\textstyle\frac{\omega}{\Omega}}\tau\right)\right)\;,\quad \tau =t\Omega\;,
\end{eqnarray}
where $z$ is the population imbalance with $z=1$ \mbox{($z=-1$)} referring to the situation with all
particles in well~1 (well~2). For low interaction $N\kappa/\Omega$ the classical phase space is regular, while for higher interaction regular and chaotic regions coexist~\cite{GuckenheimerHolmes83}.

On the $N$-particle quantum level, if all atoms occupy the single-particle state characterized by population imbalance
\begin{align}
z&= \cos^2(\theta/2) - \sin^2(\theta/2)\nonumber \\
&= \cos(\theta)
\label{eq:z_theta}
\end{align}
and relative phase~$\phi$, this leads to the wave function
\begin{eqnarray}
\label{eq:ACS}
\left|\theta,\phi\right>&=& \sum_{n=0}^N 
\left({N}\atop{n}\right)^{1/2}
\cos^{n}(\theta/2)
                 \sin^{N-n}(\theta/2)
                 \nonumber\\
                 &\times& e^{i(N-n)\phi}| n, N-n \rangle\;.
\end{eqnarray}
Here, $n$ ($N-n$) denotes the number of particles in the left (right) well. These bimodal phase-states are sometimes referred to as atomic coherent states~(ACSs)~\cite{MandelWolf95}.

Note that for finite $N$ these are in general not orthogonal,
\begin{equation}\label{eq:ACS2}
|\langle \theta^{\phantom{'}},\phi^{\phantom{'}} | \theta',\phi' \rangle |^2 > 0,\ N<\infty;
\end{equation}
while, say, $|0,N\rangle$ and  $|N,0\rangle$ are orthogonal, the scalar product of any of these two wave functions with other ACSs~(\ref{eq:ACS}) is non-zero.

The ACSs are overcomplete, to project on them we can use~\cite{MandelWolf95}
\begin{equation}
\hat{\mathbf{1}} = \frac{N+1}{4\pi}\int_0^{\pi}d\theta \sin(\theta)\int_0^{2\pi} |\theta,\phi \rangle \langle \theta,\phi |;
\end{equation}
for a given wave function $|\psi\rangle$ we can thus have the probability distribution

\begin{equation}
\label{eq:Husimi}
p_{\theta, \phi} d\Omega = \frac{N+1}{4\pi}|\langle \psi|\theta,\phi \rangle|^2 \sin(\theta)d\theta d\phi.
\end{equation}
This probability distribution is normalized to one with $0\le \theta \le \pi$ and $0\le \phi < 2\pi$.

\section{A characteristic timescale on which  $N$-particle physics deviates from mean-field for weak interactions~\label{sec:Tmf}}
One approach to explain parts of the behavior of quantum systems is to average over mean-field solutions, so called truncated Wigner methods~\cite{SinatraEtAl2002,Chuchem10,BieniasEtAl2011,SimonStrunz12,KhripkovEtAl13,GertjerenkenEtAl2013}. For a BEC in a double well, the Husimi-distribution~(\ref{eq:Husimi}) can be used to average over mean-field solutions (Refs.~\cite{UtermannEtAl94,StrzysEtAl08} and references therein). Without tilt ($\mu_0=0$) and driving  ($\mu_1=0$), the mean-field dynamics is known analytically (see, e.g., Ref.~\cite{HolthausStenholm01} and references therein). 

If the BEC initially is in one well, for low enough interactions the particles oscillate between both wells. For non-zero interactions, many-particle interactions lead to a collapse of this oscillation (which will, in a true quantum-mechanical situation eventually be followed by revivals, cf.~\cite{HolthausStenholm01,Ziegler2011,SimonStrunz12}). In this section, we derive an analytic expression for the timescale on which this collapse takes place by using the Husimi-distribution (\ref{eq:Husimi}) to mimic the apparent damping in the $N$-particle behavior.

If all particles initially are in the state $|N,0\rangle$, the Husimi-distribution becomes $\propto \sin(\theta)\cos(\theta/2)^{2N}= 2\sin(\theta/2)\cos(\theta/2)^{2N+1}$. For large $N$, this can only be non-zero for very small $\theta$, leading to the probability distribution:
\begin{equation}
\label{eq:Husimi2}
\widetilde{p}_{\theta, \phi}d\theta d\phi \simeq \frac{2N+1}{8\pi}\exp\left(-\frac{2N+1}8\theta^2\right)\theta d\theta d\phi.
\end{equation}
{To simplify the following calculations, this probability distribution is normalized to one with $0\le \theta < \infty$ and $0\le \phi < 2\pi$; contributions from angles with $\theta > \pi$ are negligible for large $N$.} Averaging over GPE-trajectories with initial conditions $\theta$ and $\phi$ with this distribution averages over states with mean-field energies $E_{\rm mf}=\cos^2(\theta)N\kappa/\Omega-\sin(\theta)\cos(\phi)\simeq (1-\theta^2)N\kappa/\Omega -\theta\cos(\phi)$.

For low interactions the system oscillates periodically, initial conditions and strength of the interactions determine amplitude and oscillation time which are known analytically~\cite{HolthausStenholm01}. For low interactions, the movement is sinusoidal~\cite{maple}. The oscillation period $1/T$~\cite{HolthausStenholm01} of those sinusoidal oscillations [$\sin(t/T)$, $\cos(t/T)$] can be expressed for low interactions and small $\theta$ as~\cite{maple}:
\begin{equation}
\frac 1T\simeq \frac 1{2\pi} + \frac{\cos(\phi)}{2\pi}\theta \frac{N\kappa}{\Omega}.
\end{equation}
The dependence of amplitude of these oscillations on the integration is a higher order effect~\cite{maple}. The next step is to average these oscillations with the Husimi-distribution~(\ref{eq:Husimi}); after integrating over $\theta$ we find~\cite{maple} damping terms:
\begin{equation}
\propto \exp\left[-2\left({\frac{N\kappa}{\Omega}}\right)^{2}{\frac {  [\cos \left( \phi \right)]^{2}}{2N+1}}{t}^{2}
\right]
\end{equation}

We thus find that $N$-particle dynamics agree with the mean-field dynamics, if $t\ll T_{\rm mf}$
\begin{equation}\label{eq:collapsetime}
T_{\rm mf} = \frac{\sqrt{2N+1}}{{\frac{N\kappa}{\Omega}}},\quad \left|\frac{N\kappa}{\Omega}\right|\ll 1\;.
\end{equation}
In the mean-field limit, $N\to \infty$ and $\kappa\to 0$ such that $N\kappa=\rm const$, the timescale on which mean-field dynamics and many-particle dynamics are expected to agree increases with $\sqrt{N}$.

\begin{figure}
\includegraphics[width = 1.0\linewidth]{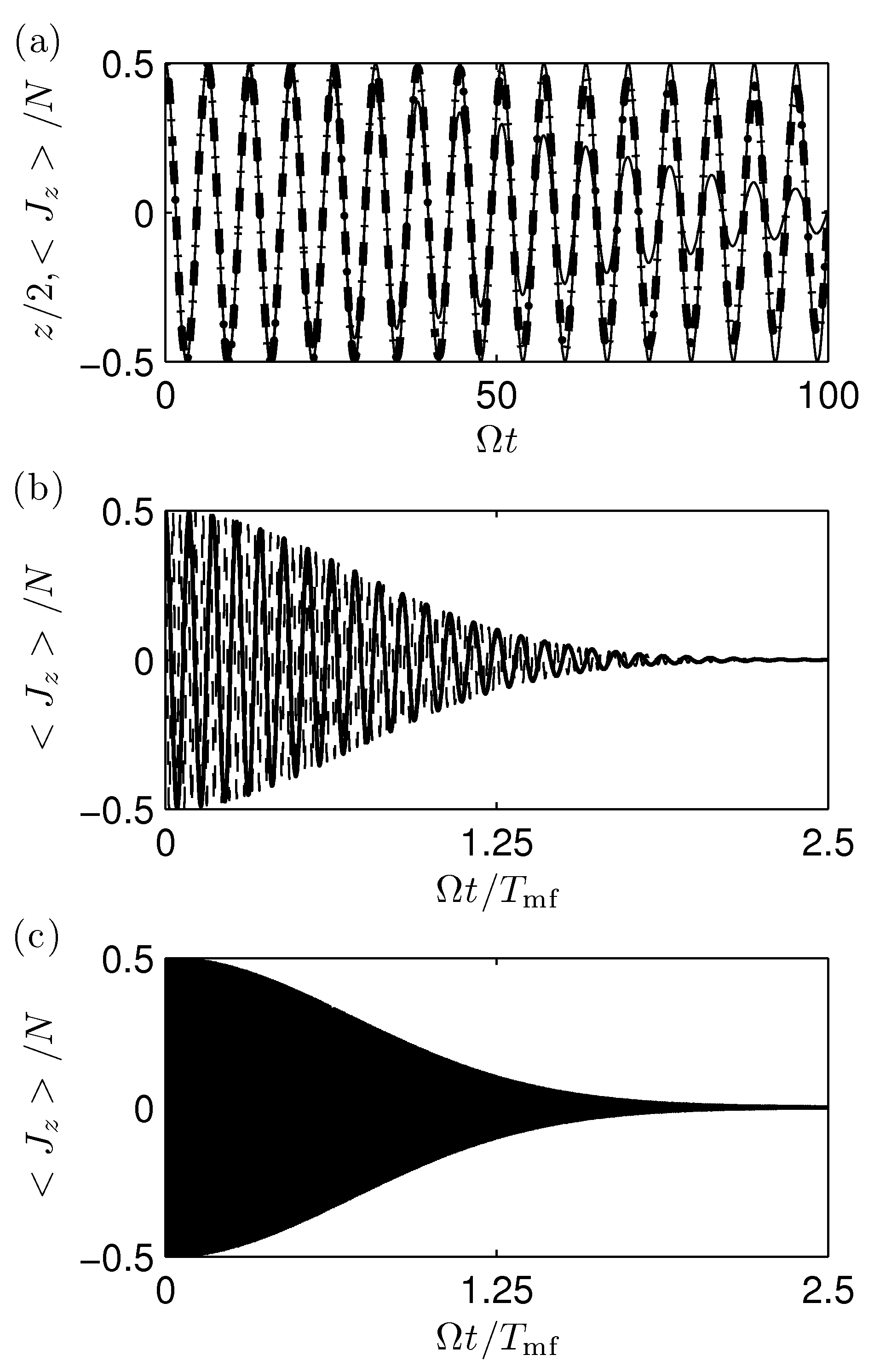}
\caption{\label{fig:undriven}Time evolution for the undriven double-well condensate with weak interaction parameter $N\kappa/\Omega = 0.2$, $\mu_0/\Omega = 0.0$ and the initial condition that all particles in the beginning are in the left well. (a) Population imbalance for GPE-solution (thin line) and $N$-particle solutions for $N = 100$ (thin line, strongly damped) and $N = 1000$ (thick dash-dotted line, weakly damped). (b) Population imbalance for $N$-particle solutions with $N = 100$ (thick line) and $N = 1000$ (thick dash-dotted line, weakly damped) with rescaled time-axis according to Eq.~(\ref{eq:collapsetime}). (c) Same as (b) for $N=10000$.}
\end{figure}

Averaging over the Husimi-distribution~(\ref{eq:Husimi}) thus predicts a damping of the oscillation, corresponding to the damping of the $N$-particle oscillations. The apparent damping is a collapse which would eventually be followed by a revival (cf.~\cite{HolthausStenholm01,Ziegler2011}). In order to explain such a behavior, extended semi-classical methods have been used~\cite{SimonStrunz12}. Averaging over classical mean-field solutions produces wrong results as soon as quantum mechanical interference plays a role.\footnote{For a Schr\"odinger cat generated via scattering a quantum bright soliton off a barrier, a truncated-Wigner calculation for the center-of-mass coordinate correctly describes the $N$-particle quantum dynamics up to the point where both parts of the wave function start to interfere again~\cite{GertjerenkenEtAl2013}. For a BEC in a strongly driven double well for which the mean-field dynamics becomes chaotic similar interferences lead to less agreement between truncated Wigner and $N$-particle quantum dynamics than for regular mean-field dynamics~\cite{WeissTeichmann08}.} We use the mean-field timescale to guide us for how long times we have to let our $N$-particle quantum dynamics run:

Figure~\ref{fig:undriven}~(a) displays numerical results for the time-evolution of the undriven double-well condensate when all the particles are initially located in the left well. While on the mean-field level the population imbalance shows full oscillations between both wells, the $N$-particle dynamics exhibit the described collapse of the oscillation, which will eventually be followed by a revival. It can be seen that the $N$-particle solutions follow the mean-field solution up to a characteristic quantum break time~\cite{TeichmannEtAl06} that increases with particle number. In Fig.~\ref{fig:undriven}~(b) and~(c) the results for the $N$-particle solutions from Fig.~\ref{fig:undriven}~(a) are displayed in rescaled time-units of $T_{\rm mf}$ according to Eq.~(\ref{eq:collapsetime}). Figure~\ref{fig:undriven}~(c) additionally shows the dynamics of the population imbalance for $N=10000$: in rescaled units the collapse takes place on the same timescale for $N=100$, $N=1000$ and $N=10000$, confirming our analytical results. The expression~$T_{\rm mf}$ from Eq.~(\ref{eq:collapsetime}) gives a good estimate for the collapse time.

\section{Approaching the mean-field limit in a periodically driven double-well}\label{sec:mfandmanybody}
\begin{figure}
\begin{center}
\includegraphics[width = 1.0\linewidth]{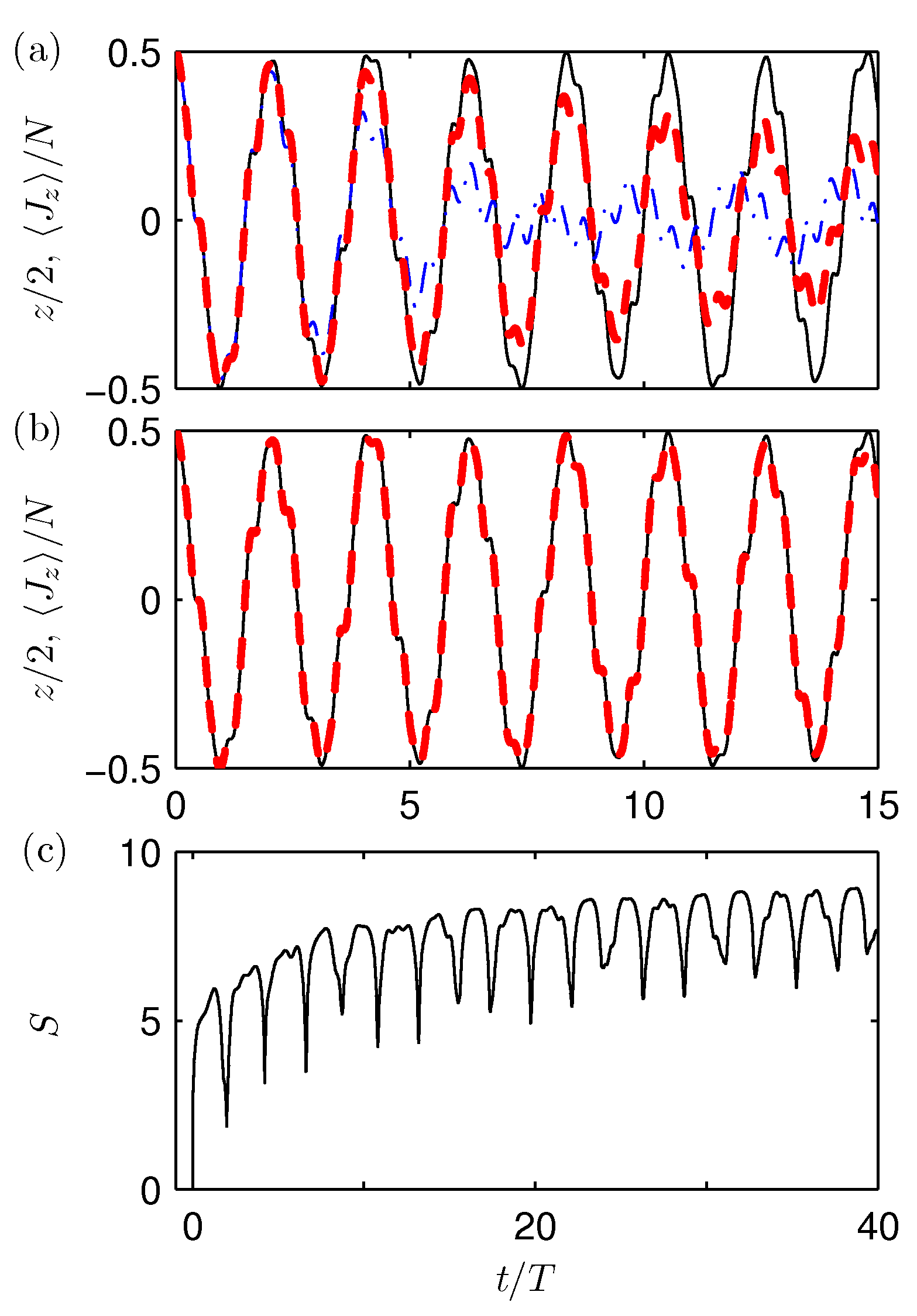}
\end{center}
\caption{\label{fig:pdiffmax}{(Color online)} (a) Time evolution for $2\mu/\omega=1.0$, $\omega/\Omega=3.0$, $\mu_0/\Omega = 0.0$ and $N\kappa/\Omega = 0.5$ with the initial condition that all particles in the beginning are in the left well. GPE-solution (thin black line), $N$-particle solutions for $N = 100$ (blue dash-dotted line), $N = 1000$ (thick red, dashed line). (b) Same as (a) for GPE-solution (thin black line) and $N=10000$ (thick, red dashed line). (c) Shannon-entropy~(\ref{eq:Shannon}) for $N=10000$. The maximum possible value $S_{\mathrm{max}}=9.21$ of the entropy is nearly reached {for some times}: $S_{\mathrm{max,num}}=8.92$. Same parameters as in (a).}
\end{figure}

In the following the model system is investigated under periodic driving. Figure~\ref{fig:pdiffmax}~(a) and~(b) displays the time-evolution of the population imbalance for particle numbers up to $N=10000$. Initially, again all particles are located in the left well. Similar to the undriven situation in Fig.~\ref{fig:undriven} the $N$-particle results exhibit the collapse of the population imbalance and the quantum break time is found to increase with particle number for the chosen parameters.

Large quantum systems are currently also actively investigated in the context of relaxation~\cite{GogolinEtAl11}. Experimentally, the relaxation to a state of maximum entropy has been observed in optical lattices~\cite{TrotzkyEtAl2012}.

For the model system investigated here, the time-evolution of the Shannon entropy~\cite{NielsenChuang00}
\begin{equation}\label{eq:Shannon}
S(t) = -\sum_{n=0}^N |a_n(t)|^2 \ln\left(|a_n(t)|^2\right)
\end{equation}
is displayed in Fig.~\ref{fig:pdiffmax}~(c) for the parameters of Fig.~\ref{fig:pdiffmax}~(a) and~(b) for $N=10000$ particles. Here, $a_n(t)$ are the coefficients in an expansion \mbox{$|\Psi(t)\rangle=\sum_{n=0}^N a_n(t)|n,N-n\rangle$} of the wave function at time $t$ over Fock states. In Fig.~\ref{fig:pdiffmax}~(c) a first rapid growing of the entropy can be observed, up to values of about $S\approx 5$. Note that this does not necessarily imply deviations of the wave function from a product state: the maximum possible value for $N=10000$ particles for the Shannon entropy~(\ref{eq:Shannon}) of an ACS~(\ref{eq:ACS}) corresponding to a mean-field state is reached for the ACS with $\vartheta = \pi/2$ and has the value $S=5.33$. In the further time evolution the value for the Shannon entropy is found to get as large as 8.92, very close to the maximum possible value of $\log(N=10000) = 9.21$ for a uniform distribution. Thus, the maximum value is nearly reached for some times. But the oscillations in the entropy in Fig.~\ref{fig:pdiffmax} indicate that in the regarded model system oscillations between the wells still take place. For systems larger than a double well, the equilibrium value would be reached for nearly all times~\cite{LindenEtAl09}.

From results as shown in figures~\ref{fig:undriven} and~\ref{fig:pdiffmax} it might be deduced that the mean-field description gets exact in the limit $N \rightarrow \infty$, $\kappa\rightarrow 0$ with $N\kappa= \mathrm{const}$: As for $N\kappa = \rm{const}$ the results of GPE and $N$-particle calculations agree for longer times $\Delta T$ with increasing particle number this could motivate the assumption: for $N\rightarrow \infty$ also $\Delta T \rightarrow \infty$. This statement has to be investigated with care. In the next section we show results that exhibit clear differences between $N$-particle dynamics and the description on the GPE-level for very large particle numbers.

\section{Beyond mean-field behavior for very large condensates}\label{sec:deviation}
\begin{figure}
\begin{center}
\includegraphics[width = 1.0\linewidth]{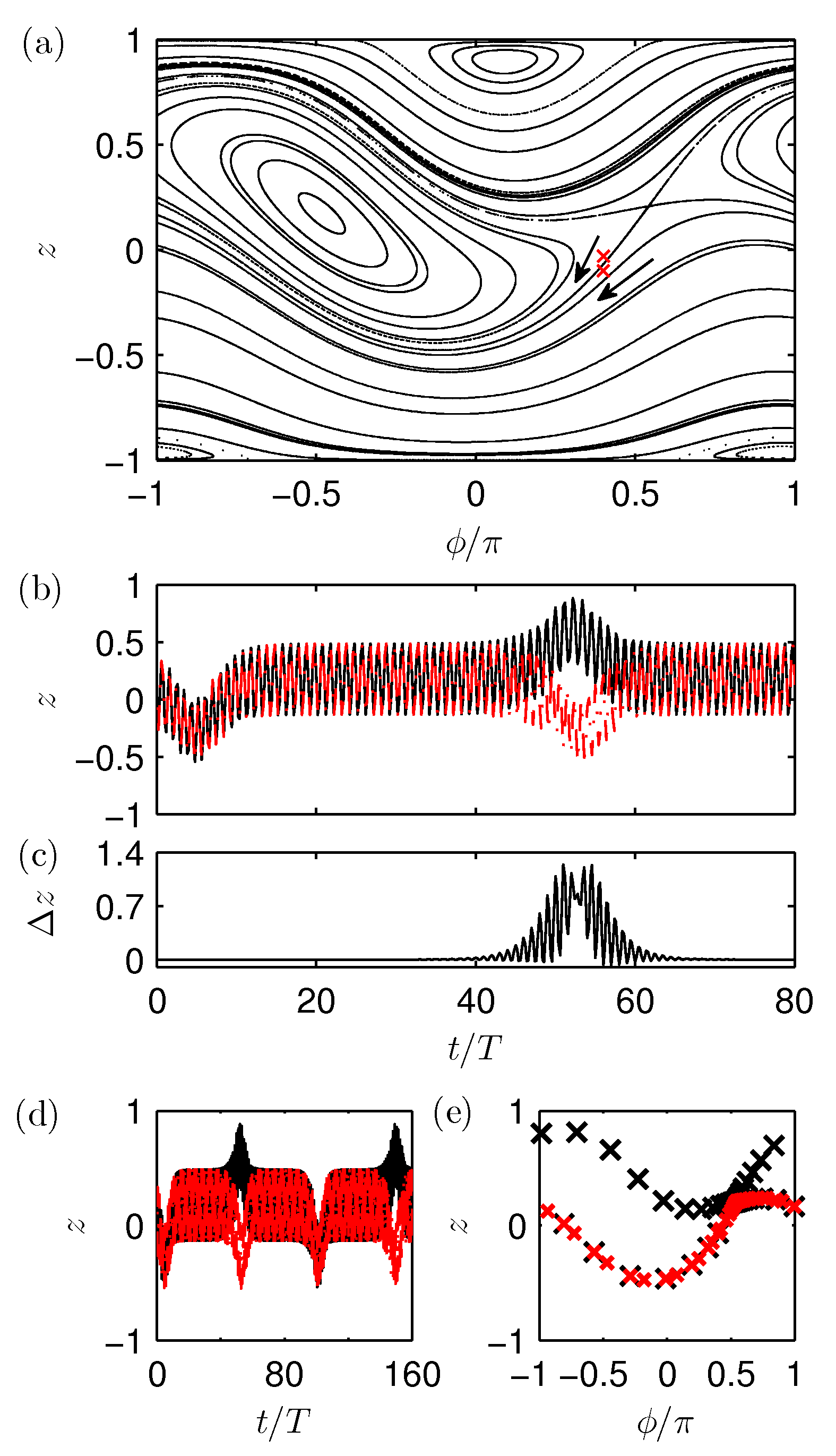}
\end{center}
\caption{\label{fig:sos}{(Color online)} (a) Poincar\'{e} surface of section for $\mu_0/\Omega = 1.5$, $N\kappa/\Omega = 0.5$, $\omega/\Omega = 3.0$ and $2\mu_1/\omega = 0.1$ with a hyperbolic fixed point at $z = 0.207$ and $\phi = 0.543\pi$. Red/gray crosses denote schematically two initial conditions with slightly different population imbalances, leading to two initial conditions on different sides of the separatrix. The black arrows indicate the direction of flow. (b)~Time evolution on the level of the GPE for two initial conditions with same phase $\phi = 0.4\pi$ but slightly different population imbalances $z_1(t=0) = -0.0689974$ (dash-dotted, red line) and $z_2(t=0) = z_1(t=0) + 10^{-7}$ (thick, black line). (c) Difference $\Delta z(t) = z_1(t)-z_2(t)$ in population imbalance. (d)~Same as~(b) for longer timescale. (e)~Mean-field trajectories depicted at integers of the period duration~$T=2\pi/\omega$ for~$0 < t/T < 80$. Small red/gray crosses: initial condition $z_1(t=0)$, large black crosses: $z_2(t=0)$.}
\end{figure}
In the following for exemplary initial conditions the time-evolution is discussed both on the mean-field and on the $N$-particle level to demonstrate deviations from (GP)-mean-field behavior for large particle numbers.

For the comparison of mean-field and $N$-particle dynamics the relation to the phase space of the corresponding classical system is of special interest: in~\cite{Chuchem10} the convergence to classicality is investigated for different initial conditions. Temporal fluctuations in the bosonic Josephson junction have also been investigated as a probe for phase space tomography~\cite{KhripkovEtAl13}. In~\cite{TeichmannEtAl06} a periodically driven double-well system with a mixed phase space was investigated and it was shown that the mean-field limit is approached rapidly with $N$ in regular regions of phase space, but that strong differences occur in chaotic regions of phase space. The relation between mean-field chaos and entanglement in periodically driven double-well condensates was investigated in~\cite{WeissTeichmann08,GertjerenkenEtAl10}: in chaotic regions of phase space the creation of entanglement is accelerated~\cite{WeissTeichmann08}. Here, we focus on very large particle numbers for initial conditions close to the separatrix in a mainly regular phase space. 

For weak driving the Poincar\'{e} surface of section corresponding to the classical Hamiltonian~(\ref{eq:mean}) is displayed in Fig.~\ref{fig:sos}~(a). The separatrix divides phase space into regions with qualitatively different types of motion: ``oscillation'' for closed trajectories around the elliptic fixed point at $z\approx0.18$ and $\phi\approx-0.47\pi$ and ``rotation''. The regular islands around the elliptic fixed points at $z=0.9$, $\phi=0.1\pi$ and $z=-0.97$, $\phi=0.95\pi$ correspond to the self-trapping regime~\cite{SmerziEtAl97,AlbiezEtAl05}.

Such a classical perspective can give important insight into $N$-particle dynamics. On the GPE-level a wave-function is characterized by the parameters $\theta$ and $\phi$, representing a point in classical phase space. As the ACSs are not orthogonal the associated $N$-particle state~(\ref{eq:ACS}) has a certain extension in phase space that gets smaller with increasing particle number $N$ and eventually vanishes in the mean-field~(\ref{eq:meanfieldlimit}). To account for these quantum mechanical uncertainties often semi-classical methods, where a phase-space distribution is propagated, are used~\cite{UtermannEtAl94,SinatraEtAl2002,StrzysEtAl08,WeissTeichmann08,Chuchem10,BieniasEtAl2011,SimonStrunz12,KhripkovEtAl13,GertjerenkenEtAl2013}. On the $N$-particle level it was demonstrated in~\cite{GertjerenkenEtAl10} that a hyperbolic fixed point acts as a generator of mesoscopic entanglement. The relation to the classical phase space has also been investigated experimentally: it was demonstrated at the example of the internal Josephson effect that a quantum mechanical many-particle system can exhibit a classical bifurcation~\cite{ZiboldEtAl10}.

Now, two initial mean-field conditions (red crosses in Fig.~\ref{fig:sos}~(a)) are chosen such that they are closely spaced but located to either side of the separatrix. Both states have equal relative phase~$\phi$ and slighty different population imbalances $z_1(t=0)=-0.0689974$ and $z_2(t=0) = z_1(t=0) + \Delta z (t=0)$ with $\Delta z(t=0) = 10^{-7}$. The black arrows in Fig.~\ref{fig:sos}~(a) indicate the direction of flow, leading to the the time-evolution of the population imbalance~$z(t)$ depicted in Fig.~\ref{fig:sos}~(b): While the mean-field trajectories initially stay closely spaced, at the hyperbolic fixed point the trajectories diverge and the different types of motion for initial conditions to either side of the separatrix become visible. The clearly distinguishable behavior around $t/T\approx 55$ is highlighted in Fig.~\ref{fig:sos}~(c) where the difference $\Delta z(t) = z_1(t)-z_2(t)$ in population imbalance is depicted. In Fig.~\ref{fig:sos}~(d) it can be seen that similar differences occur repeatedly also at later times. In Fig.~\ref{fig:sos}~(e) the time-evolution of the two mean-field trajectories is visualized in dependence of population imbalance~$z$ and relative phase~$\phi$ for times~$0 < t/T < 80$. As data points are always depicted at integer multiples of the period duration it can be seen that the motion is slowed down close to the hyperbolic fixed point. This explains why clearly visible differences between both trajectories occur only in a short time interval, when the trajectories move away from the hyperbolic fixed point. As both trajectories return to the hyperbolic fixed point at different times in the long-time behavior clearly visible differences occur more often.

On the $N$-particle level the unitarity of the time evolution operator $U\left(t-t_0\right)$ implies that the scalar product of two $N$-particle states $|\Psi_1\left( t\right) \rangle$ and $|\Psi_2\left( t\right) \rangle$ is the same at all times $t$:
\begin{eqnarray}\label{eq:unitarityscal}
\langle \Psi_1\left(t\right) | \Psi_2\left( t\right) \rangle & =& \langle \Psi_1\left(t_0\right) |U^{\dagger}\left(t-t_0\right) U\left(t-t_0\right) | \Psi_2\left( t_0\right) \rangle \nonumber \\ &=& \langle \Psi_1\left(t_0\right) | \Psi_2\left( t_0\right) \rangle.
\end{eqnarray}
Thus, the scalar product of initially very close ACSs~(\ref{eq:ACS}) stays close to one for all times $t$. It only deviates from one for very large~$N$. For the initial ACSs~(\ref{eq:ACS}) corresponding to the mean-field initial conditions in Fig.~\ref{fig:sos} the value of the scalar product still is 0.99 for 
\begin{equation}
N=10^{12}.
\label{eq:huge}
\end{equation}
 For the presented situation this implies intuitively that the full $N$-particle dynamics cannot be captured by the (GP)-mean-field approximation even for very large particle numbers.

\begin{figure*}
\includegraphics[width = 1.0\linewidth]{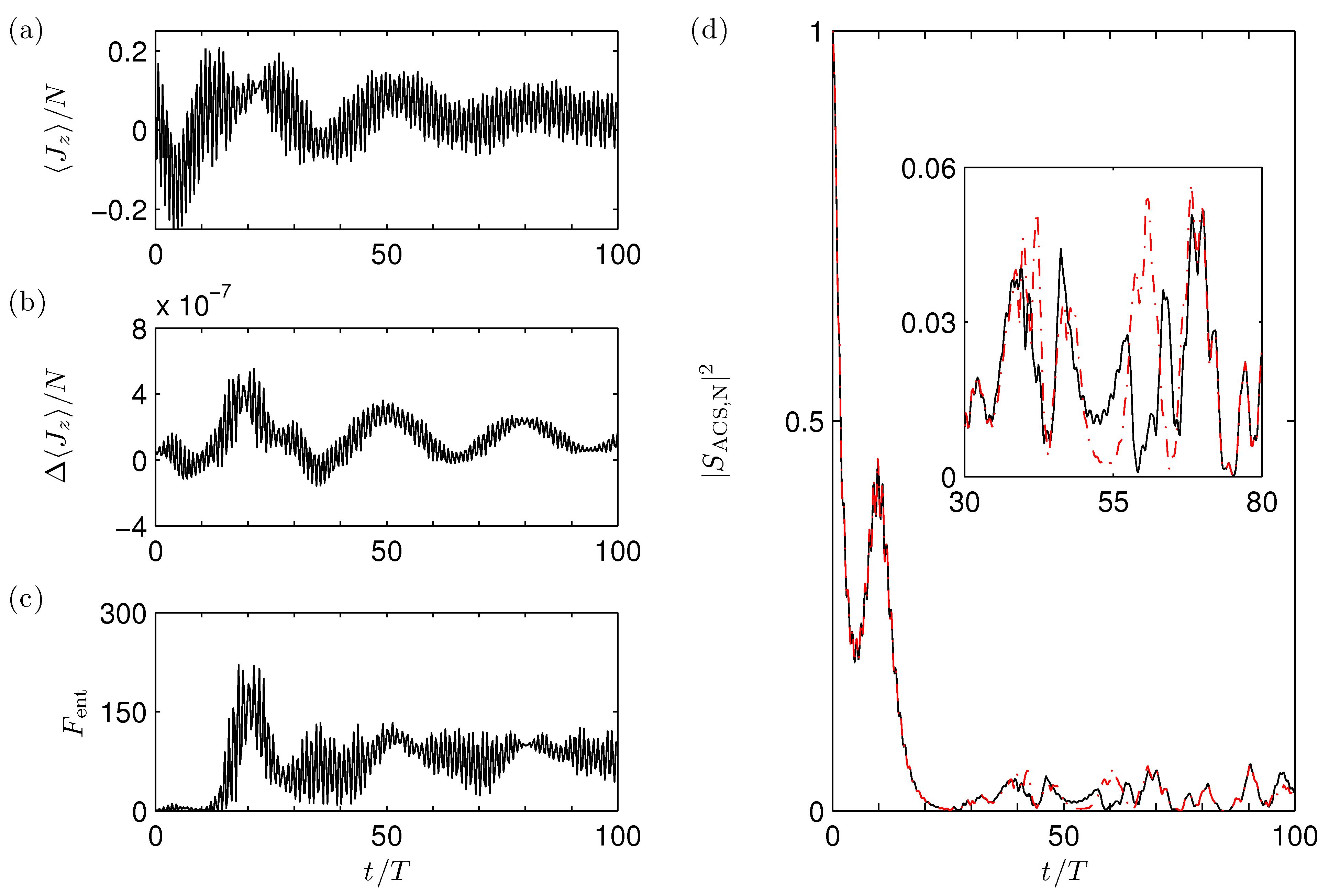}
\caption{\label{fig:scalprod}{(Color online)} Time evolution on the $N$-particle level for~$N=1000$ and the same parameters as in Fig.~\ref{fig:sos}. (a)~Population imbalance~$\langle J_z \rangle/N$ between the two potential wells for one of the initial conditions. (b)~Difference~$\Delta \langle J_z \rangle/N$ in~$\langle J_z \rangle/N$ for the two initial conditions. (c)~Entanglement flag~(\ref{eq:QFI2}) for one of the initial conditions, indicating entanglement with~$F_{\mathrm{ent}}>1$ for times~$t/T>15$. (d)~Scalar product of mean-field and $N$-particle state at each point of time for both initial conditions (black, solid and red, dash-dotted line). The time-evolution on the $N$-particle level is initialized with the ACS corresponding to the initial mean-field state. To calculate the scalar product at each point of time of the mean-field time-evolution the corresponding ACS is computed.}
\end{figure*}

For 1000 particles and the parameters from Fig.~\ref{fig:sos} this point is illustrated in Fig.~\ref{fig:scalprod}~(d), where the time evolution of the absolute square of the scalar product
\begin{equation}\label{eq:scalprod}
S_{\mathrm{ACS,N}}(t)= \langle \vartheta(t),\varphi (t)|\Psi (t) \rangle 
\end{equation}
is shown. Here, $|\vartheta(t),\varphi (t)\rangle$ is the ACS~(\ref{eq:ACS}) associated at each point of time with the time-evolved mean-field state and $|\Psi (t)\rangle$ denotes the time-evolved $N-$particle state when the time-evolution is initialized with the ACS~(\ref{eq:ACS}) corresponding to the initial mean-field state. The time evolution of the absolute square of the scalar product~(\ref{eq:scalprod}) is displayed for both initial states and a rapid drop to very low values is observed, confirming the statement that the $N$-particle dynamics cannot follow the dynamics of the ACSs~(\ref{eq:ACS}). Differences in the two curves correspond to the differences between the two mean-field states in Fig.~\ref{fig:sos}~(b). The data in Fig.~\ref{fig:scalprod}~(d) can be understood in the following way: on the $N$-particle level the initial wave function has a certain extension, as the ACSs are not orthogonal. This wave function is then torn apart when crossing the hyperbolic fixed point, stretching along the separatrix. This behavior is nearly the same for both initial states, if they are close enough to each other. The mean-field states  [respectively the resulting ACSs~(\ref{eq:ACS})] -- lying on either side of the separatrix and thus corresponding to distinct behavior -- cannot show this behavior.

Additionally, the corresponding expectation value~$\langle J_z \rangle/N$ with the population imbalance~(\ref{eq:angularmom}) is investigated on the $N$-particle level. It can be proved analytically, independent of the model used, that quantum mechanical wave functions $|\Psi_1\rangle$ and $|\Psi_2\rangle$ (normalized to one) which are similar in the sense of
\begin{equation}
\label{eq:delta}
|\langle \Psi_1(0)|\Psi_2 (0)\rangle|^2 = 1- {\delta} \;\;\mathrm{with}\;\;{\delta} \ll 1.
\end{equation} lead to similar expectation values $\langle A \rangle$ for operators like~$\hat{A}=\hat{J}_z/N$ (Appendix~\ref{app:similar}). For the difference
\begin{equation}
\Delta \langle J_z \rangle \equiv \langle \Psi_2(t) |\hat{J_z}| \Psi_2(t)\rangle - \langle \Psi_1(t) |\hat{J_z}| \Psi_1(t)\rangle
\end{equation}
of expectation values for the population imbalance $J_z$ we find [Eqs.~(\ref{eq:Adiff}) and (\ref{eq:CforJ})]:\footnote{\label{page:footnote}The prove of Eq.~(\ref{eq:error}), which can be found in the appendix, uses properties of the operator $\hat{J}_z/N$ that are shared with other (but not all) operators. For a particle in a one-dimensional quantum mechanical box, $0\le x\le L$, the expectation values $\langle \psi_1|x|\psi_1\rangle$ and $\langle \psi_2|x|\psi_2\rangle$ would also lie close together if the scalar product of both wave functions satisfied Eq.~(\ref{eq:delta}). However, if the size of the box goes to infinity, the differences can become arbitrarily large.}
\begin{equation}
\label{eq:error}
\frac{|\Delta \langle J_z \rangle|}N \le 2\left[ \sqrt{{\delta}} + \mathcal{O}\left({\delta}\right)\right] \;\;\mathrm{for}\;\;{\delta} \ll 1.
\end{equation}
This statement is also visible in the numerics for the investigation of Fig.~\ref{fig:pdiffmax}: in Fig.~\ref{fig:scalprod}~(a) numerical results for the expectation value $\langle J_z \rangle$ of the population imbalance on the $N$-particle level are shown for one of the initial conditions. In~(b) the difference $\Delta \langle J_z \rangle$ in $\langle J_z \rangle$ for both initial conditions is displayed, which is on the order of $10^{-6}$ for $N=1000$. Thus, even on the level of expectation values, the difference between both states remains small. The numeric investigations of Fig.~\ref{fig:scalprod} are a graphic illustration of the more general result~(\ref{eq:error}). This result, proved in the appendix, remains true for the huge BEC of Eq.~(\ref{eq:huge}). Thus, on the $N$-particle level the differences between the dynamics of the two initial conditions will be very small, in particular very much smaller than the difference on the mean-field level depicted in Fig.~\ref{fig:sos}.

The deviations between mean-field dynamics and $N$-particle quantum dynamics have been described to indicate emergence of entanglement on the $N$-particle level~\cite{WeissTeichmann08}, which we can also observe in Fig.~\ref{fig:scalprod}. Here, we use as a signature for entanglement the quantum Fisher information~\cite{PezzeSmerzi2009,footnoteQFIneu3} for the relative phase between the condensates in the two potential wells. For pure states it reads
\begin{equation}
\label{eq:QFI}
 F_{\rm QFI}\equiv 16{\left(\langle J_z^2\rangle-\langle J_z\rangle^2\right)}\;,
\end{equation}
with the experimentally measurable~\cite{EsteveEtAl08} variance $\langle J_z^2\rangle-\langle J_z\rangle^2$ of the population imbalance. An entanglement flag is then given by
\begin{equation}
\label{eq:QFI2}
 F_{\rm ent} >1,\,\quad F_{\rm ent}\equiv  \frac{F_{\rm QFI}}N.
\end{equation}
This is a sufficient condition for particle-entanglement and identifies those entangled states that are useful to overcome classical phase sensitivity in interferometric measurements~\cite{PezzeSmerzi2009}.

For the presented situation the entanglement flag~(\ref{eq:QFI2}) is displayed in~\ref{fig:scalprod}~(c). For times $t/T \gtrsim 15$ it takes on values $F_{\mathrm{ent}}> 1$, indicating entanglement.

\section{Conclusion}\label{sec:conclusion}
We have presented an example for which $N$-particle quantum dynamics remains different from mean-field (Gross-Pitaevskii) dynamics up to particle numbers of the order of $10^{12}$. This number is significantly higher than for the largest condensates in experiments with ultracold atoms (consisting of about $10^{9}$ atoms  of spin-polarized hydrogen~\cite{FriedEtAl98}). To do this, we identify that two distinct initial conditions near the separatrix lead to a mean-field behavior that cannot be reproduced by even the largest BECs (let alone the small BECs in double wells investigated, e.g., in the experiment~\cite{AlbiezEtAl05}, cf.~\cite{ZiboldEtAl10}).

By combining analytic and numeric investigations, we have shown that the differences between mean-field dynamics and $N$-particle quantum dynamics would be visible on the level of expectations values on experimentally realistic short timescales. For initial conditions far away from the separatrix, the timescale on which mean-field and $N$-particle quantum physics agree would be larger than the lifetime of the BEC for such huge BECs [Eq.~(\ref{eq:collapsetime})].

While we have chosen the specific model system of a double-well potential in two-mode approximation, deductions for more general systems can be drawn (cf.~footnote on page~\pageref{page:footnote}).

\section{Acknowledgements}
We thank S.~Arlinghaus, S.~A.~Gardiner, M.~Hiller and M.~Holthaus for discussions. B.G. thanks C.~S.~Adams and S.~A.~Gardiner for hospitality at the University of Durham and acknowledges funding by the 'Studienstiftung des deutschen Volkes' and the 'Heinz Neum\"uller Stiftung'. B.G. acknowledges support from the Deutsche Forschungsgemeinschaft through grant no. HO 1771/6-2.

\appendix

\section{\label{app:similar}Similar initial wave functions can lead to similar expectation values at all times}

If two initial wave functions are different from each other, we can write their overlap as:
\begin{equation}
\label{eq:over}
|\langle\psi_1(0) |\psi_2(0)\rangle|^2= 1-\delta, 
\end{equation}
with
\begin{equation}
0<\delta\le 1\;.
\end{equation}
As shown in Eq.~(\ref{eq:unitarityscal}) quantum mechanics yields that the scalar product of both functions remains constant for all times.

A similar statement does, however, not necessarily apply to expectation values: if, say, a small part of a spatial single-particle  wave function is moved very far away this has hardly any effect on a scalar product but can have large effects on calculating the expectation value of the position.

In Sec.~\ref{sub:In} we show that the operators $\hat{A}=\hat{J}_{x,y,z}/N$ defined in Eqs.~(\ref{eq:angularmom}) are bounded in the sense:
\begin{align}
\label{eq:bounded}
{\left|\langle\psi|\hat{A}|\widetilde{\psi}\rangle\right|}\le C\;,\quad 0<C< \infty\;,
\end{align}
where the constant $C$ is independent of $N$ and for all wave functions normalized to 1. For such cases, we show in Sec.~\ref{sub:prove} that the difference of the expectation values remains small for all times:
\begin{equation}
|\langle\psi_2(t)|\hat{A}|\psi_{2}(t)\rangle - \langle\psi_1(t)|\hat{A}|\psi_{1}(t)\rangle|\le 2C\left[\sqrt{\delta(1-\delta)} +\delta\right]
\end{equation}
if the scalar product of both wave functions is close to one [cf.~Eq.~(\ref{eq:over})]

\subsection{\label{sub:In}The operators $\hat{J}_{x,y,z}/N$  are bounded in the sense of Eq.~(\ref{eq:bounded})}

For the model discussed in this paper, all wave functions can be expressed in the Fock-bases, i.e:
\begin{equation}
|\psi\rangle \equiv \sum_{n=0}^Nb_n|n,N-n\rangle
\end{equation}
and 
\begin{equation}
|\widetilde{\psi}\rangle \equiv \sum_{n=0}^Nc_n|n,N-n\rangle
\end{equation}
with
\begin{align}
\label{eq:norm1}
\sum_{n=0}^N|b_n|^2 = 1,\quad
\sum_{n=0}^N|c_n|^2 = 1.
\end{align}
We now can show:
\begin{align}
\frac 1N\left|\langle\psi|{\hat{a}^{\dag}_1\hat{a}^{\phantom{\dag}}_1}|\widetilde{\psi}\rangle\right| &= \frac 1N\left|\sum_{n=0}^N b_n^*nc_n\right|\nonumber\\
&\le \frac 1N\sum_{n=0}^N  \left|b_n^*\right|n \left|c_n\right|\nonumber\\
&\le\frac 1N N\sum_{n=0}^N  \left|b_n\right| \left|c_n\right|\;.
\label{eq:smaller}
\end{align}
Because of  the inequality 
\begin{equation}
\label{eq:ineq}
|b||c|\le (|b|^2+|c|^2)/2
\end{equation} 
[which is valid because of $(|b|-|c|)^2>0$], the remaining sum in Eq.~(\ref{eq:smaller}) is smaller than:
\begin{align}
\sum_{n=0}^N  \left|b_n\right| \left|c_n\right| &\le  \sum_{n=0}^N  \frac12\left(\left|b_n\right|^2+ \left|c_n\right|^2\right)\nonumber\\
&=1\;;
\label{eq:smaller2}
\end{align}
for the last step we have used Eqs.~(\ref{eq:norm1}).

Thus, we have proved the first of the following four inequalities:
\begin{align}
\label{eq:a11}
\frac 1N\left|\langle\psi|{\hat{a}^{\dag}_1\hat{a}^{\phantom{\dag}}_1}|\widetilde{\psi}\rangle\right|&\le 1\\
\label{eq:a22}
\frac 1N\left|\langle\psi|{\hat{a}^{\dag}_2\hat{a}^{\phantom{\dag}}_2}|\widetilde{\psi}\rangle\right|&\le 1\\
\label{eq:a21}
\frac 1N\left|\langle\psi|{\hat{a}^{\dag}_2\hat{a}^{\phantom{\dag}}_1}|\widetilde{\psi}\rangle\right|&\le 1\\
\label{eq:a12}
\frac 1N\left|\langle\psi|{\hat{a}^{\dag}_1\hat{a}^{\phantom{\dag}}_2}|\widetilde{\psi}\rangle\right|&\le 1\;.
\end{align}
The prove of Eq.~(\ref{eq:a22}) goes analogously to the above prove of Eq.~(\ref{eq:a11}). To show  Eq.~(\ref{eq:a21}) we can use:
\begin{align}
\frac 1N\left|\langle\psi|{\hat{a}^{\dag}_2\hat{a}^{\phantom{\dag}}_1}|\widetilde{\psi}\rangle\right| &= \frac 1N\left|\sum_{n=1}^N b_n^*\sqrt{n}\sqrt{N-n+1}c_{n-1}\right|\nonumber\\
&\le \frac 1N\sum_{n=1}^N  \left|b_n^*\right|N \left|c_{n-1}\right|\nonumber\\
&\le  \sum_{n=1}^N  \frac12\left(\left|b_n\right|^2+ \left|c_{n-1}\right|^2\right)\nonumber\\
&\le 1\;;
\label{eq:smaller3}
\end{align}
and for the prove of Eq.~(\ref{eq:a12}):
\begin{align}
\frac 1N\left|\langle\psi|{\hat{a}^{\dag}_1\hat{a}^{\phantom{\dag}}_2}|\widetilde{\psi}\rangle\right| &= \frac 1N\left|\sum_{n=0}^{N-1} b_n^*\sqrt{n+1}\sqrt{N-n}c_{n+1}\right|\nonumber\\
&\le \frac 1N\sum_{n=0}^{N-1}  \left|b_n^*\right|N \left|c_{n+1}\right|\nonumber\\
&\le  \sum_{n=0}^{N-1}  \frac12\left(\left|b_n\right|^2+ \left|c_{n+1}\right|^2\right)\nonumber\\
&\le 1\;,
\label{eq:smaller4}
\end{align}
which again uses Eq.~(\ref{eq:ineq}).
Because of $|x\pm y|\le |x| + |y|$, this also proves the inequalities:
\begin{equation}
\frac 1N\left|\langle\psi|\hat{J}_{\zeta}|\widetilde{\psi}\rangle\right| \le 1\;;\quad \zeta\in\{x,y,z\}\;.
\label{eq:CforJ}
\end{equation}

\subsection{\label{sub:prove}Prove that for operators bounded in the sense of Eq.~(\ref{eq:bounded}), similar wave functions have similar expectation values}
In the following, we use that the wave functions $|\psi_{1}(t)\rangle$ and $|\psi_{2}(t)\rangle$ are normalized:
\begin{align}
\langle \psi_{1}(t)|\psi_{1}(t)\rangle &= 1\\
\langle \psi_{2}(t)|\psi_{2}(t)\rangle &= 1\;.
\end{align}
For any such functions for which Eq.~(\ref{eq:over}) is valid, we can express $|\psi_{2}(t)\rangle$ as:
\begin{align}
|\psi_{2}(t)\rangle = \sqrt{1-\delta}e^{i\alpha(t)}&|\psi_{1}(t)\rangle \nonumber\\\
+\sqrt{\delta}&|\psi_{\perp}(t)\rangle
\end{align}
where $\alpha(t)$ is a real number and
\begin{align}
\langle\psi_1(t)|\psi_{\perp}(t)\rangle &= 0\\
\langle\psi_{\perp}(t)|\psi_{\perp}(t)\rangle &= 1\;.
\end{align}

For all operators $\hat{A}$ we have:
\begin{align}
\langle\psi_2(t)|\hat{A}|\psi_{2}(t)\rangle = \quad& \langle\psi_1(t)|\hat{A}|\psi_{1}(t)\rangle(1-\delta)\nonumber\\
+&\langle\psi_{\perp}(t)|\hat{A}|\psi_{1}(t)\rangle\sqrt{\delta(1-\delta)}e^{i\alpha(t)}\nonumber\\
+&\langle\psi_{1}(t)|\hat{A}|\psi_{\perp}(t)\rangle\sqrt{\delta(1-\delta)}e^{-i\alpha(t)}\nonumber\\
+&\langle\psi_{\perp}(t)|\hat{A}|\psi_{\perp}(t)\rangle\delta
\end{align}
Defining the difference of the expectation values as:
\begin{equation}
\Delta\langle A\rangle \equiv \langle\psi_2(t)|\hat{A}|\psi_{2}(t)\rangle - \langle\psi_1(t)|\hat{A}|\psi_{1}(t)\rangle
\end{equation}
yields
\begin{align}
\left|\Delta\langle A\rangle\right|\le\quad& \left|\langle\psi_1(t)|\hat{A}|\psi_{1}(t)\rangle\right|\delta\nonumber\\
+& \left|\langle\psi_{\perp}(t)|\hat{A}|\psi_{1}(t)\rangle\right|\sqrt{\delta(1-\delta)}\nonumber\\
+& \left|\langle\psi_{1}(t)|\hat{A}|\psi_{\perp}(t)\rangle\right|\sqrt{\delta(1-\delta)}\nonumber\\
+& \left|\langle\psi_{\perp}(t)|\hat{A}|\psi_{\perp}(t)\rangle\right|\delta\;.
\end{align}

If $\hat{A}$ is bounded in the sense that for all $\psi$ and $\widetilde{\psi}$
with $\langle\psi|\psi\rangle =1$ and $\langle\widetilde{\psi}|\widetilde{\psi}\rangle =1$ and for all $N$ the following inequality~(\ref{eq:bounded}),
\begin{align}
\nonumber
{\left|\langle\psi|\hat{A}|\widetilde{\psi}\rangle\right|}\le C\;,\quad 0<C< \infty\;,
\end{align}
where $C$ is independent of $N$, is true, we thus have:
\begin{align}
\left|\Delta\langle A\rangle\right|&\le 2C\left[\sqrt{\delta(1-\delta)} +\delta\right].
\end{align}
which becomes
\begin{align}
\left|\Delta\langle A\rangle\right|&\le 2C\left[\sqrt{\delta} +{\mathcal O}(\delta)\right]\quad {\rm for}\quad\delta\ll 1\; .
\label{eq:Adiff}
\end{align}


\begin{thebibliography}{52}
\bibitem[{foo()}]{footnoteQFIneu3}
\bibinfo{note}{Here the observable $\vec{J}_{\vec{n}}$ from
  \cite{PezzeSmerzi2009} is taken to be the particle number difference
  $n_{12}$. Hence, a quantum Fisher information relative to the relative phase
  is calculated.}
\end{thebibliography}
\end{document}